\documentclass[letterpaper, 10 pt, conference]{ieeeconf}  

\IEEEoverridecommandlockouts                              

\usepackage{cite}
\usepackage{amsmath,amssymb,amsfonts}
\usepackage{algorithmic}
\usepackage{graphicx}
\usepackage{textcomp}
\usepackage{caption}
\usepackage{subcaption}
\usepackage{siunitx}
\usepackage{float}
\usepackage{physics}
\usepackage{multirow}
\usepackage{hyperref}
\usepackage{array}
 \usepackage{booktabs}

\renewcommand{\baselinestretch}{0.98}

\title{\LARGE \bf
Mixed Integer Linear Programming for Active Contact Selection in Deep Brain Stimulation
}

\author{Anna Franziska Frigge$^{1}$ and Alexander Medvedev$^{2}$
\thanks{*This work is funded by the Swedish Research Council via Grant 2020-02901 to the project ``Patient-specific dynamical modeling and optimization of deep brain stimulation" within The EU Joint Programme – Neurodegenerative Disease Research.}
\thanks{$^{1}$Anna Frigge is with the Division of Systems and Control, Department of Information Technology,
        Uppsala University, Uppsala, Sweden
        {\tt\small anna.frigge@it.uu.se}}%
\thanks{$^{2}$Alexander Medvedev is with the Division of Systems and Control, Department of Information Technology,
        Uppsala University, Uppsala, Sweden
        {\tt\small alexander.medvedev@it.uu.se}}%
}

\begin{document}

\maketitle
\thispagestyle{empty}
\pagestyle{empty}

\begin{abstract}

Deep brain stimulation (DBS) programming remains a complex and time-consuming process, requiring manual selection of stimulation parameters to achieve therapeutic effects while minimizing adverse side-effects. This study explores mathematical optimization for DBS programming, using functional subdivisions of the subthalamic nucleus (STN) to define the desired activation profile. A Mixed Integer Linear Programming (MILP) framework is presented  allowing for dissimilar current distribution across active contacts. MILP is compared to a Linear Programming (LP) approach in terms of computational efficiency and activation accuracy. Results from ten Parkinson’s disease patients treated with DBS show that while MILP better matches the predefined stimulation target activation profile, LP solutions more closely resemble clinically applied settings, suggesting the profile may not fully capture clinically relevant patterns. Additionally, MILP’s limitations are discussed, including its reliance on precisely defined target regions and its computational burden for larger target sets. 
\newline

\indent \textit{Clinical relevance}—  A framework that facilitates and expedites the  DBS programming procedure by predicting optimal current distributions is established.
\end{abstract}

\section{Introduction}

Deep Brain Stimulation (DBS) is an invasive treatment applied in neurology and psychiatry. DBS consists of chronically delivering electrical pulses of certain amplitude, length, and frequency to a stimulation target in the brain via a surgically implanted lead. The stimulation target is selected individually and on the basis of the treated medical condition. DBS has been an FDA-approved treatment for Parkinson’s Disease (PD) since 2002, offering substantial benefits in managing motor symptoms and improving the quality of life for patients~\cite{FDA2002,Honey2017}.  The impulses are produced by a battery-driven Implantable Pulse Generator (IPG), typically placed below the clavicle. The primary stimulation targets in PD are the subthalamic nucleus (STN) and globus pallidus internus  (GPi)~\cite{Honey2017}. Individualization of the DBS treatment aims at achieving maximal symptom alleviation with minimal stimulation side effects and is carried out by selecting suitable lead contacts, or combinations of thereof, as well as the parameters of the pulse train. This procedure is referred to as DBS programming.

Despite significant advances in lead design, surgical techniques, and clinical guidelines, the DBS programming process remains largely manual and time-consuming. Programming involves identifying optimal stimulation parameters for individual patients, a task traditionally handled through a laborious trial-and-error approach that requires substantial clinician involvement. This process is further complicated  and hindered by the lack of automation and standardized objective symptom quantification tools.
Current DBS programming practices primarily rely on subjective evaluations by clinicians, often measured by scores such as the Unified Parkinson’s Disease Rating Scale (UPDRS)~\cite{UPDRS2003}. Emerging technologies, including smartphones, smartwatches, gait sensors, and video-based assessments, have begun to offer objective data for evaluating patient responses~\cite{Olsson2020, Hadley2021,Zhang2024,Sarapata2023}. However, integrating these technologies into a comprehensive assessment of patient state remains challenging.

Computational modeling and image-guided tools have demonstrated the potential to enhance and expedite the DBS programming. By visualizing the implanted lead relative to key brain structures, these tools have been shown to reduce the initial programming duration by providing clinicians with a more informed starting point for parameter selection~\cite{Aldred2023}. However, fully automated optimization frameworks capable of handling the complex parameter space and patient-specific constraints are still under development. The introduction of increasingly sophisticated lead designs offers extensive means of shaping the stimulation field but as also complicates parameter selection in clinical practice. While state-of-the-art leads currently have eight contacts, the recent introduction of a 16-contact lead (Vercise\textsuperscript{{\texttrademark}} Cartesia\textsuperscript{{\texttrademark}} X Directional Lead) comes with a drastic increase in the number of degrees of freedom in DBS programming. Exploring all possible configurations in a clinical setting is infeasible and emphasizes the need for  tools automating the DBS parameter selection.  

As mentioned before, the goal of DBS programming is readily formulated as an optimization problem, although in clinical terms of symptoms and side effects. Several mathematical optimization approaches for electrical brain stimulation problems, e.g. DBS, have been proposed, each with specific strengths and limitations~\cite{Frigge2024PP}. Linear programming (LP) methods offer simplicity and computational efficiency but may lack flexibility. Nonlinear optimization approaches provide greater coherence at the cost of increased complexity and computational demand. Additionally, modern IPGs  offer greater flexibility in stimulation  current distribution across active contacts, which adds another level of complexity to DBS programming and has been neglected in previous approaches of optimization-based parameter selection for DBS~\cite{Frigge2024PP,Cubo2019}.

Notably, the stimulation amplitude at a contact takes real values whereas the selection of active contacts in a DBS lead is a distinctively discrete problem. A Mixed Integer Linear Programming (MILP) optimization framework is introduced in \cite{Abouelseoud2018} for a general electrical brain stimulation (EBS) scenario. 
 In this work, we adapt and apply the MILP approach specifically to the DBS programming scenario and compare its performance to a simpler LP-based optimization method. The main goal is to extend the functionality of open-source DBS optimization pipeline TuneS \cite{Frigge2024PP} with more versatile algorithms. Novelty of the paper is in the fact that MILP has not been considered in a DBS programming setting before.

 The main contributions of this paper are as follows:
\begin{itemize}
    \item An optimization framework accounting  for dissimilar current values across the set of active contacts is proposed.
    \item The limitations of MILP in DBS programming are discerned. In particular, the optimization technique relies on precisely defined target and constraint regions, and poorly scales with increasing number of target points.
    \item LP and MILP algorithms are compared in terms of computational burden and their ability to achieve a desired activation profile.
    \item LP and MILP algorithms are further compared in their ability to match clinically used stimulation settings.
\end{itemize}

The rest of the paper is organized as follows. First, the considered patient cohort and the clinical data used in the study  are presented. Then, the patient-specific mathematical model of DBS underlying the optimization approach is described. Further, the optimization algorithms employed, i.e. LP and MILP, and their use in the DBS programming are explained. The results obtained by optimization of the  models constructed for the individual leads and their comparison to the clinical stimulation settings constitute the main part of the paper, followed up by a brief discussion and conclusions.

\section{Patient cohort}

\begin{figure}
    \centering
    \begin{subfigure}{0.50\linewidth}
        \centering
        \includegraphics[width=\linewidth]{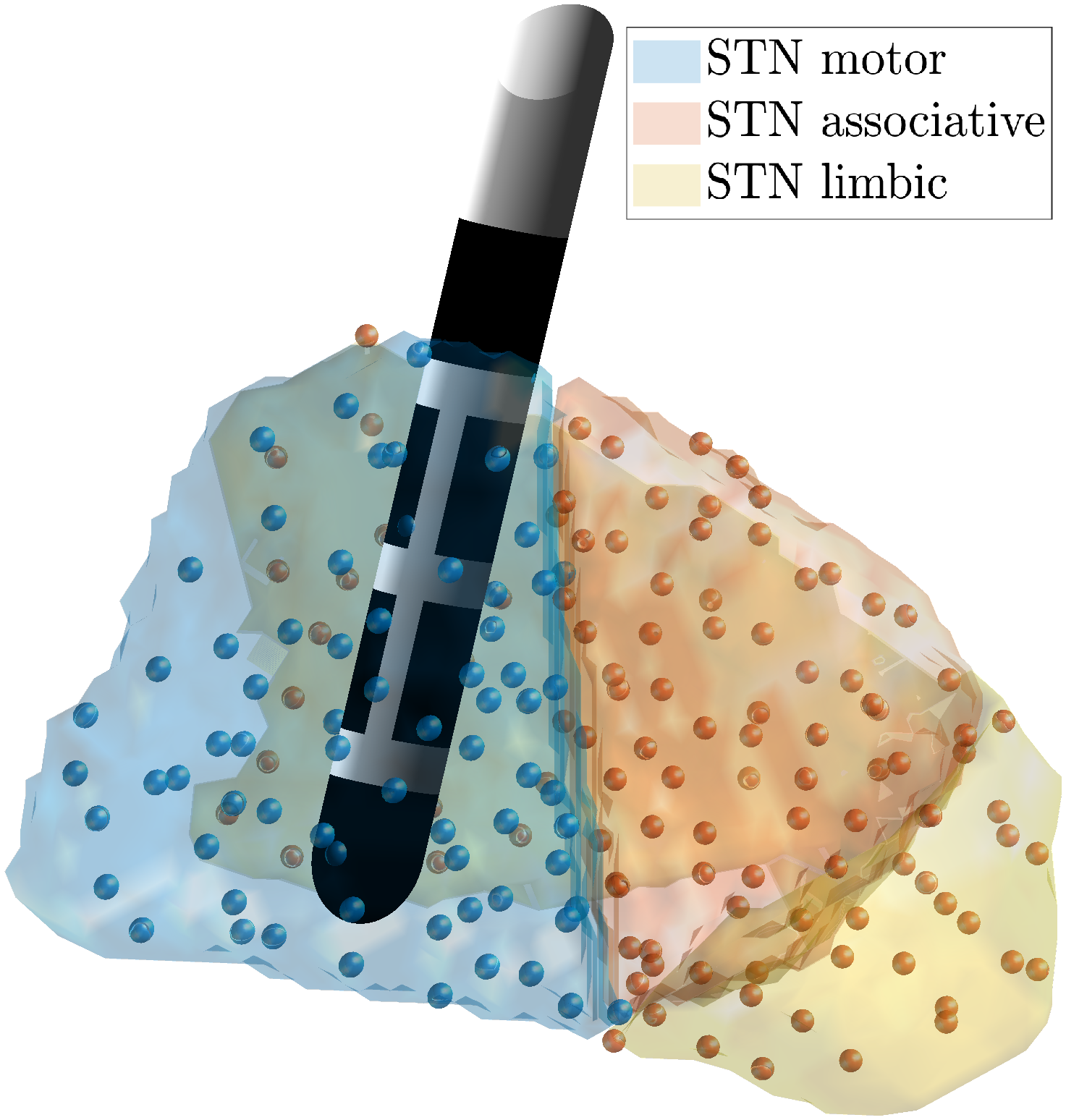}
        \caption{}
        \label{fig:DBS_STN_points}
    \end{subfigure} \hspace{0.75cm}
    \begin{subfigure}{0.27\linewidth}
        \centering
        \includegraphics[width=\linewidth]{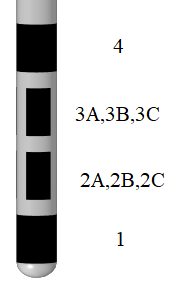}
        \caption{}
        \label{fig:Abbot_lead}
    \end{subfigure}
    \caption{(a) An eight-contact Boston Scientific Vercise Cartesia\textsuperscript{\texttrademark} Directional lead relative to the functional subdivisions of the STN~\cite{Ewert2018} along with target (blue) and constraint (orange) points. The original set of points were taken from an atlas~\cite{Ewert2018} and downsampled using a voxel filter with a voxel length of $\SI{0.95}{mm}$.
    (b) The Abbot's Medical Infinity\textsuperscript{\texttrademark} Directional lead with the contact nomenclature used in this study. The upper and lower contact rows are ring contacts, while the two middle rows consist of three segmented contacts (A,B,C) each.}
    \label{fig:main_figure}
\end{figure}

\begin{figure}
    \centering
    \includegraphics[width=0.8\linewidth]{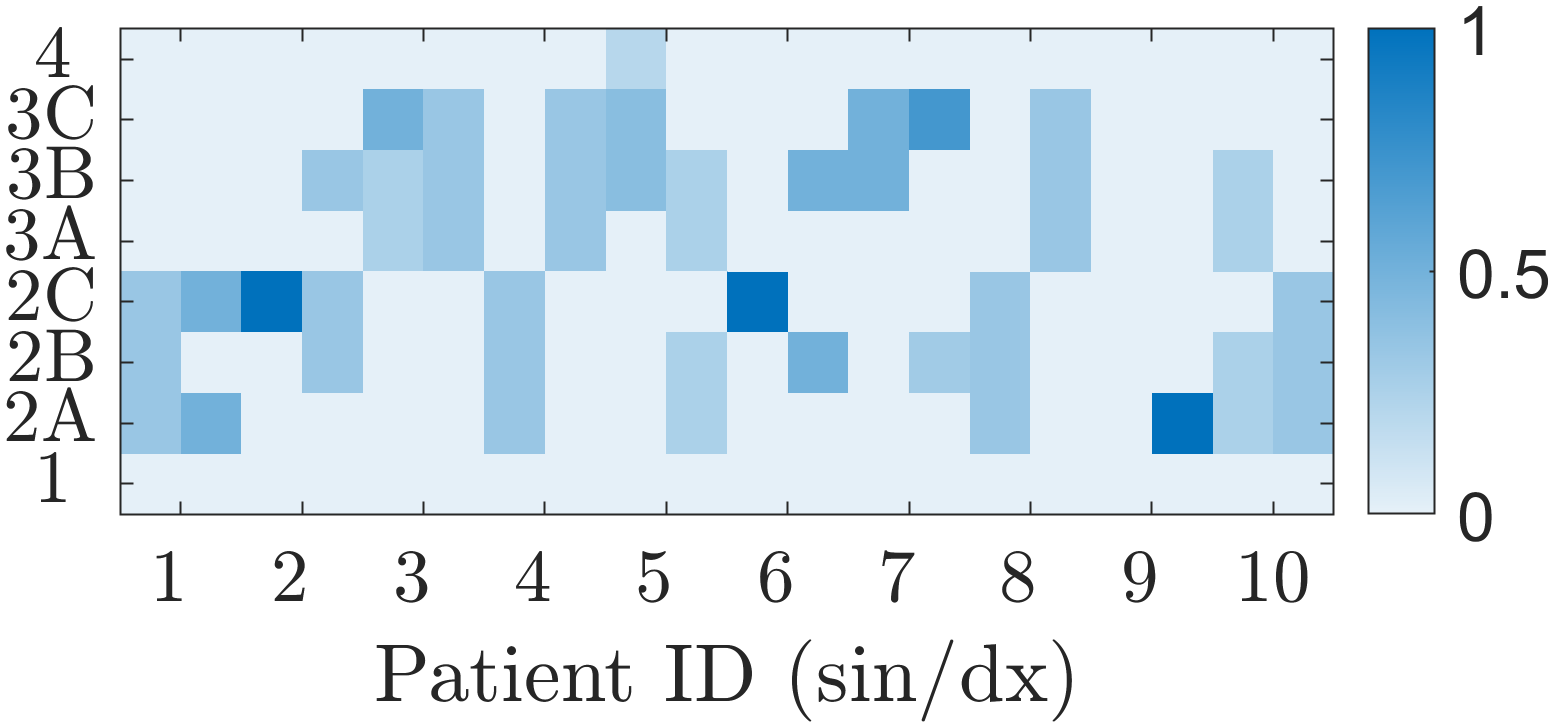}
    \caption{The distribution of current through all contacts under clinical settings. Values range between zero (light blue) and one (dark blue). Patient ID -- $x$-axis.  Active contact -- $y$-axis. }
    \label{fig:ClinicalSettings}
\end{figure}

This study includes a cohort of ten PD patients treated at Uppsala University Hospital\footnote{The study was approved by the Swedish Ethics Review Authority, Registration number 
2019–05718, and all participants gave informed written consent prior to the beginning of the study.}. Nine of these patients received bilateral DBS, resulting in a total of 19 implanted leads.
All patients were implanted with state-of-the-art, eight-contact directional leads. The cohort was evenly divided between two lead models: five patients received the Boston Scientific Vercise Cartesia\textsuperscript{\texttrademark} Directional lead, whereas the other five were implanted with the Abbot's Medical Infinity\textsuperscript{\texttrademark} Directional lead. The lead designs along with the contact labels used in this paper are illustrated in Fig.~\ref{fig:DBS_STN_points} and Fig.~\ref{fig:Abbot_lead}. In all cases, the intended surgical target was the STN. The clinical stimulation settings, including active contacts and their respective current distributions, are summarized in Fig.~\ref{fig:ClinicalSettings}. Throughout this study, the two brain hemispheres are referred to as \textit{sin} (Latin: \textit{sinister}, left) and \textit{dx} (Latin: \textit{dexter}, right), respectively.
Preprocessing of pre- and postoperative imaging data, as well as reconstruction of lead coordinates, was performed using Lead-DBS v2.6\cite{Horn2019}, as described in~\cite{Frigge2024PP}, which analyzed the same dataset.

\section{Mathematical modeling}

In DBS modeling, a common and practical approach to estimate the volume of tissue activated (VTA) involves identifying regions where the electric field norm $E$ exceeds a given threshold, $E_{\mathrm{th}}$. A point $k$ is considered activated (or excited), if $E_k \geq E_{\mathrm{th}}$. VTAs offer a simple yet effective approximation of the spread of an external stimulus in neural tissue, bypassing the complexity of  consistent neuronal orientations definition in gray matter structures, such as the STN, which is the focus of this study.  

The spread of the electric field in neural tissue surrounding the DBS lead is simulated using a Finite Element Method (FEM) model. A static approximation of the electric field is computed by solving the following partial differential equation (PDE) that describes the distribution of  the electric potential $u$ in three dimensions
\begin{equation}
    \nabla \cdot (\sigma \nabla u) = 0,
    \label{eq:pde_static}
\end{equation}
where $\sigma$ denotes conductivity. This static formulation assumes negligible capacitive and inductive effects. Non-active contact surfaces are assigned floating boundary conditions. In this study, the PDE is solved  using COMSOL Multiphysics\textsuperscript{\textregistered} 5.6  with Livelink\textsuperscript{\texttrademark} for MATLAB. 
Further details about the model and the workflow of building individualized models of stimulation spread in DBS are provided in~\cite{Frigge2024PP}.

For a DBS lead with $N$ contacts, \eqref{eq:pde_static} is solved $N$ times, each time applying a unit stimulus of $\SI{1}{mA}$  to a single contact while all others contacts remain inactive. 
For a set of points $\Omega$ with $N_{\Omega}$ locations relevant to the problem-setup, the electric field norm $E_{kp}$ for a unit stimulus applied to contact $p$ at each location $k=1,\dots,N_{\Omega}$ is interpolated to construct the transfer matrix $\mathbf{T}$, defined as
\begin{equation}
\mathbf{T} =
  \begin{bmatrix}
    E_{11}       & E_{12} & E_{13} & \dots & E_{1N} \\
    E_{21}       & E_{22} & E_{23} & \dots & E_{2N} \\
    \vdots       & \vdots & \vdots &       & \vdots\\
    E_{N_{\mathrm{\Omega}}1}       & E_{N_{\mathrm{\Omega}}2} & E_{N_{\mathrm{\Omega}}3} & \dots & E_{N_{\mathrm{\Omega}}N}
\end{bmatrix}.
\end{equation}

The linearity of electrical model \eqref{eq:pde_static} allows scaling the unit stimulus solution to any desired current amplitude. When multiple contacts belonging to the set $p=1,...,N$ are simultaneously active, the cumulative electric field norm can be computed by superimposing individual solutions. This is expressed as
\begin{equation}
    \mathbf{y} =  \mathbf{T} \mathbf{u},
    \label{eq:superposition}
\end{equation}
where $\mathbf{u} = [u_1,...,u_{N}]$ is the vector of current amplitudes applied to each contact, and $\mathbf{y}$ is the resulting electric field norm at all relevant locations.

Switching a contact from inactive to active alters the boundary conditions in \eqref{eq:pde_static}, making exact superposition of individual fields strictly speaking incorrect. Specifically, the superimposed solutions include the influence of induced currents at other contacts, which should not be present when a contact is actively stimulating. As a result, neglecting the effect of induced currents is expected to slightly overestimate the electric field norm, in particular near the DBS lead.  

In this paper, unipolar DBS configurations, meaning one or multiple DBS lead contacts acting as the cathode, while the IPG is the anode, are assumed. Bipolar configurations are only inaccurately represented by the electric field norm activating function~\cite{Duffley2019,Frigge2024Mathmod} and are therefore not considered in this paper.

Provided that the electric field norm is calculated in the same system of coordinates, the overlap of two VTAs $X$ and $Y$, can be evaluated using the Dice-S{\o}rensen coefficient, given by 
\begin{equation}
    D = \frac{|X\cap Y|}{|X|+|Y|},
\end{equation}
where $|\cdot|$ denotes the cardinality of a set.

\section{Optimization}
\subsection{Optimization objective}
The DBS programming scenario can be interpreted as an optimization problem, where the objective is to activate specific target regions associated with therapeutic benefits while minimizing stimulation in areas linked to undesirable side effects. In the case of STN DBS, the STN can be subdivided into functional regions, i.e. the motor, limbic, and associative STN~\cite{Ewert2018}, cf. Fig~\ref{fig:DBS_STN_points}.  In this paper, it is assumed that stimulation of the motor region is beneficial for alleviating common motor symptoms in PD, whereas activation of the limbic and associative regions is undesirable due to the risk of stimulation-induced side effects. Recent discussions, however, have emphasized the importance of DBS in modulating activity in white matter tracts connected to the STN~\cite{Hollunder2024}. While this study leaves white matter fiber activation out of scope, future work could incorporate these tracts with their known orientations to explore which targets are most associated with clinical outcomes.

With respect to the discussion above, DBS programming is formalized as a procedure seeking activation in a set of target points  $\Omega_{\mathrm{t}}$, while avoiding activation in a set of constraint points $\Omega_{\mathrm{c}}$.  This implies $\Omega_{\mathrm{t}} \cap \Omega_{\mathrm{c}}=\emptyset$, where $|\Omega_{\mathrm{t}}|=N_{\mathrm{t}}$ and $|\Omega_{\mathrm{c}}|=N_{\mathrm{c}}$.  For cases with multiple targets or constraints, the union of disjoint sets defines the target and constraint volumes, respectively.
In this work, the clinical DBS programming problem is simplified to focus exclusively on optimizing contact selection and current distribution among active contacts to achieve the desired activation profile, without accounting for dynamic stimulation parameters. Additionally, all target and constraint points are treated as equally favorable or unfavorable for a positive (or adverse) stimulation effect, notwithstanding of their position.

Fig.~\ref{fig:DBS_STN_points} illustrates the atlas-based target and constraint regions~\cite{Ewert2018} along with the DBS lead for one patient. The original set of target and constraint points was downsampled using a voxel grid filter, where all points within a voxel were approximated by their centroid. To evaluate the computational performance of the two optimization algorithms presented in the following sections, the number of points was varied using different voxel sizes.

All computations were performed in MATLAB, and the optimization problems were solved using the Gurobi Optimizer.

\subsection{Safety constraints}

To ensure the safety of the proposed stimulation settings, constraints are imposed on the solution of the optimization problem. The following safety constraints are defined:
\begin{equation}
    \begin{aligned}
        0 \leq I_p \leq I_{\mathrm{max}}, &\quad \forall  p\in \lbrack 1, N \rbrack ,\\
    I_{\mathrm{tot}}\triangleq &\sum_{p=1}^N I_p \leq I_{\mathrm{max}}. 
    \end{aligned}
    \label{eq:safetyConstraints}
\end{equation}
Here, $I_p$ represents the current delivered through each individual contact, while $I_{\mathrm{tot}}$ denotes the total current amplitude across all contacts. These constraints ensure that no individual contact exceeds the maximum allowable current and that the combined total current remains within safe limits. In clinical practice, these limits may vary depending on patient-specific sensitivity to stimulation. In this paper, the current limits were set to $I_p=\SI{5}{mA}$ and $I_{\mathrm{tot}}=\SI{8}{mA}$. As only unipolar DBS configurations are treated, all current amplitudes are assumed positive. Unlike the approach in~\cite{Abouelseoud2018}, no explicit limits are imposed on the amplitude of the induced field in target areas to prevent ``overstimulation". 

\subsection{Linear programming}
The DBS programming can be cast as a linear programming problem, where the goal is to maximize the electric field norm at all target points while ensuring that it remains below a specified threshold, $E_{\mathrm{th,off}}$, at all or a given percentage of constraint points. This can be expressed as
\begin{equation}
\begin{aligned}
        \mathbf{u^*} = &\arg\max_{\mathbf{u}} \quad \sum_i \mathbf{T}_i \mathbf{u}, \quad \forall i \in \Omega_\mathrm{t} \\
        \text{s.t.} \quad& \mathbf{T}_j \mathbf{u} \leq E_{\mathrm{th,c}}, \quad \text{for } \theta \text{ }\% \text{ of points } j \in \Omega_\mathrm{c},
\end{aligned}
    \label{eq:LP}
\end{equation}
where $\mathbf{T}_i \mathbf{u}$ represents the electric field norm at each target location $i$  resulting from the superposition of fields induced by the stimulation amplitudes $\mathbf{u}$ on all contacts. Similarly, $\mathbf{T}_j \mathbf{u}$ denotes the electric field norm at each constraint location $j$. The parameter $\theta$ allows for relaxation of the constraints and is intended to account for patient-specific sensitivity to stimulation.

This linear programming formulation is adapted from~\cite{Frigge2024PP}, where the direct superposition of fields, as described in \eqref{eq:superposition}, was not utilized due to its technical inaccuracy. Instead, that work considered only uniform current distribution across all contacts and optimized for pre-defined contact combinations. By incorporating the superposition of fields from individual contacts in this model, it is possible to directly optimize the proportional distribution of currents across all contacts, albeit potential small overestimation of the superimposed VTA.



\subsection{Mixed Integer Linear Programming}

A general approach to solving electrical stimulation problems using MILP was introduced in \cite{Abouelseoud2018}. In this work, the formulation was adapted to the context of DBS by using VTA, rather than induced currents, to quantify stimulation effects. Additionally, since only unipolar stimulation is considered, constraints on the reference electrode (anode) or the inclusion of negative current amplitudes become redundant.


Under these considerations, the MILP formulation  can be rewritten for DBS as
\begin{equation}
\begin{aligned}
    \mathbf{u^*} = &\arg\min_{\mathbf{u},\mathbf{d}} \quad  \left(  \frac{1}{N_\mathrm{t}}\sum_{i=1}^{N_\mathrm{t}} d_i + \frac{1}{N_\mathrm{c}}\sum_{j=1}^{N_\mathrm{c}} d_{j} \right), \\
    \text{s.t.} \quad & \mathbf{T_i} \mathbf{u} + L d_i \geq E_{\mathrm{th,t}}, \quad \forall i \in \Omega_\mathrm{t},\\
    & \mathbf{T_j} \mathbf{u} - L d_j \leq E_{\mathrm{th,c}}, \quad \forall j \in \Omega_\mathrm{c},\\
            &d_i,d_j \in \{0,1\},
    \end{aligned}
    \label{eq:MILP}
\end{equation}
where $d_i$ and $d_j$ are binary dummy variables and $L$ is a larger number that allows for the relaxation of the constraint at a given target or constraint point. To allow for sufficient constraint relaxation, $L$ should be chosen significantly larger than $E_{\mathrm{th,t}}$, but too large $L$ may result in computational inefficiency. 

A solution derived from~\eqref{eq:MILP} directly allows to determine the number of targets points and constraint points that are activated and therefore either fulfill or violate the optimization objective.


\subsection{Performance assessment measures}

Suggested contact configurations can be assessed with regard to their ability to achieve the desired activation profile. The degree of similarity between the achieved activation profile, $\mathbf{P_{\mathrm{act}}}$, and the desired activation profile,  $\mathbf{P_{\mathrm{des}}}$, is quantified using the inconsistency metric $\beta$ defined as~\cite{Abouelseoud2018}

\begin{equation}
    \beta = 0.5\left(\frac{N_{\mathrm{t,n}}}{N_\mathrm{t}}+\frac{N_{\mathrm{c,a}}}{N_\mathrm{c}}\right).
    \label{eq:inconsistency}
\end{equation}
Here, $N_{\mathrm{t,n}}$ denotes the number of non-activated target points (missed desired activations), while $N_{\mathrm{c,a}}$ represents the number of activated constraint points (undesired activations).
The overall objective should be to minimize inconsistency between a desired activation profile and achieved activation profile. 
 For the MILP formulation in~\eqref{eq:MILP}, the inconsistency can be directly computed from the optimal solution vector. In contrast, for the LP formulation in~\eqref{eq:LP}, the superimposed field at the target and constraint points is computed from $\mathbf{Tu^*}$. The values of $N_{\mathrm{t,n}}$ and $N_{\mathrm{c,a}}$ are then determined by applying the respective thresholds for target and constraint points.

Cohort-level results can be represented as a matrix, where each column corresponds to the current distribution for a single DBS lead. To quantitatively compare these distributions across the cohort, the differences between LP and MILP solutions, as well as between clinical settings (cf. Fig.~\ref{fig:ClinicalSettings}) and the respective optimization results are evaluated. The Frobenius norm computing the element-wise deviation between two matrices, $\mathbf{A}$ and $\mathbf{B}$, is employed to quantify these differences:
\begin{equation}
    ||\mathbf{A}-\mathbf{B}||_\mathrm{F} = \sqrt{\sum_{n,m} (a_{n,m}-b_{n,m})^2}.
    \label{eq:Frobenius}
\end{equation}

\section{Results}
All results were computed in the Montreal Neurological Institute (MNI) space~\cite{Fonov2009}.

\subsection{Accuracy of superimposed VTAs}
\begin{figure}
    \centering
    \includegraphics[width=0.8\linewidth]{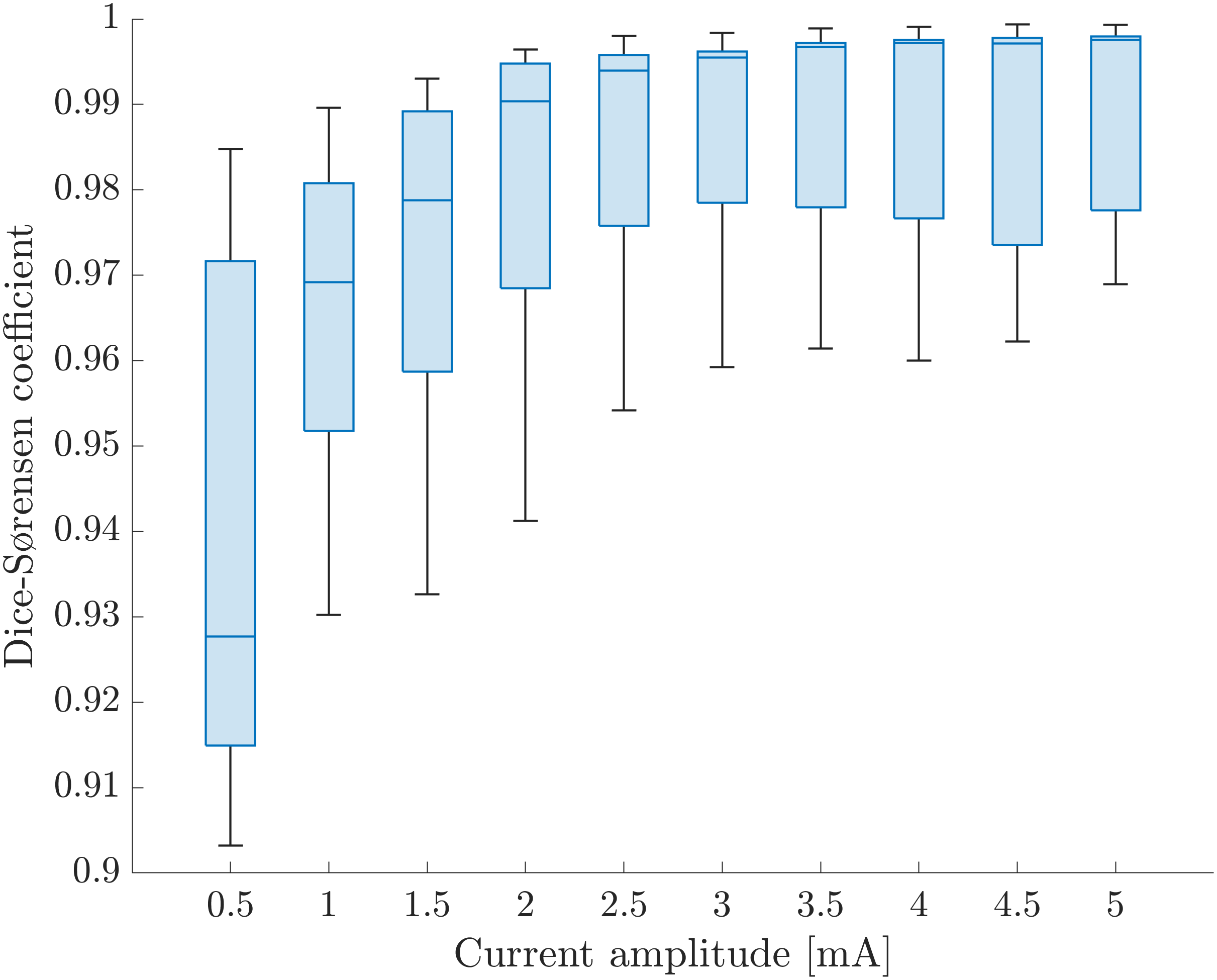}
    \caption{Dice-S{\o}rensen coefficients comparing VTAs directly computed from the solution of the COMSOL model with VTAs reconstructed via superposition of individual contact solutions. The boxplot statistics were computed from a total of $23$ contact configurations across amplitudes from $\SI{0.5}{mA}$ to $\SI{5}{mA}$, including adjacent segmented contacts, three-segment rings, ring-segment pairings, and vertically misaligned combinations.}
    \label{fig:Dice_VTAs}
\end{figure}
Fig.~\ref{fig:Dice_VTAs} shows the Dice-S{\o}rensen coefficients comparing VTAs directly computed from the COMSOL model with VTAs reconstructed by superimposing solutions from individual contacts. A total of 23 contact combinations were evaluated for amplitudes ranging from $\SI{0.5}{mA}$ to $\SI{5}{mA}$.
These combinations include horizontally and vertically adjacent segmented contacts, three-segment ring configurations, ring contacts paired with single segmented contacts, and vertically misaligned segmented contact arrangements. All reported Dice-S{\o}rensen coefficients fall in the range between $0.9$ and $1$, indicating strong overall agreement between the VTAs. The coefficients increase with higher amplitudes, likely due to the reduced influence of floating  contacts on the overall size of the VTA at larger stimulation strengths.

\subsection{Linear programming results}
\begin{figure}
    \centering
    \includegraphics[width=\linewidth]{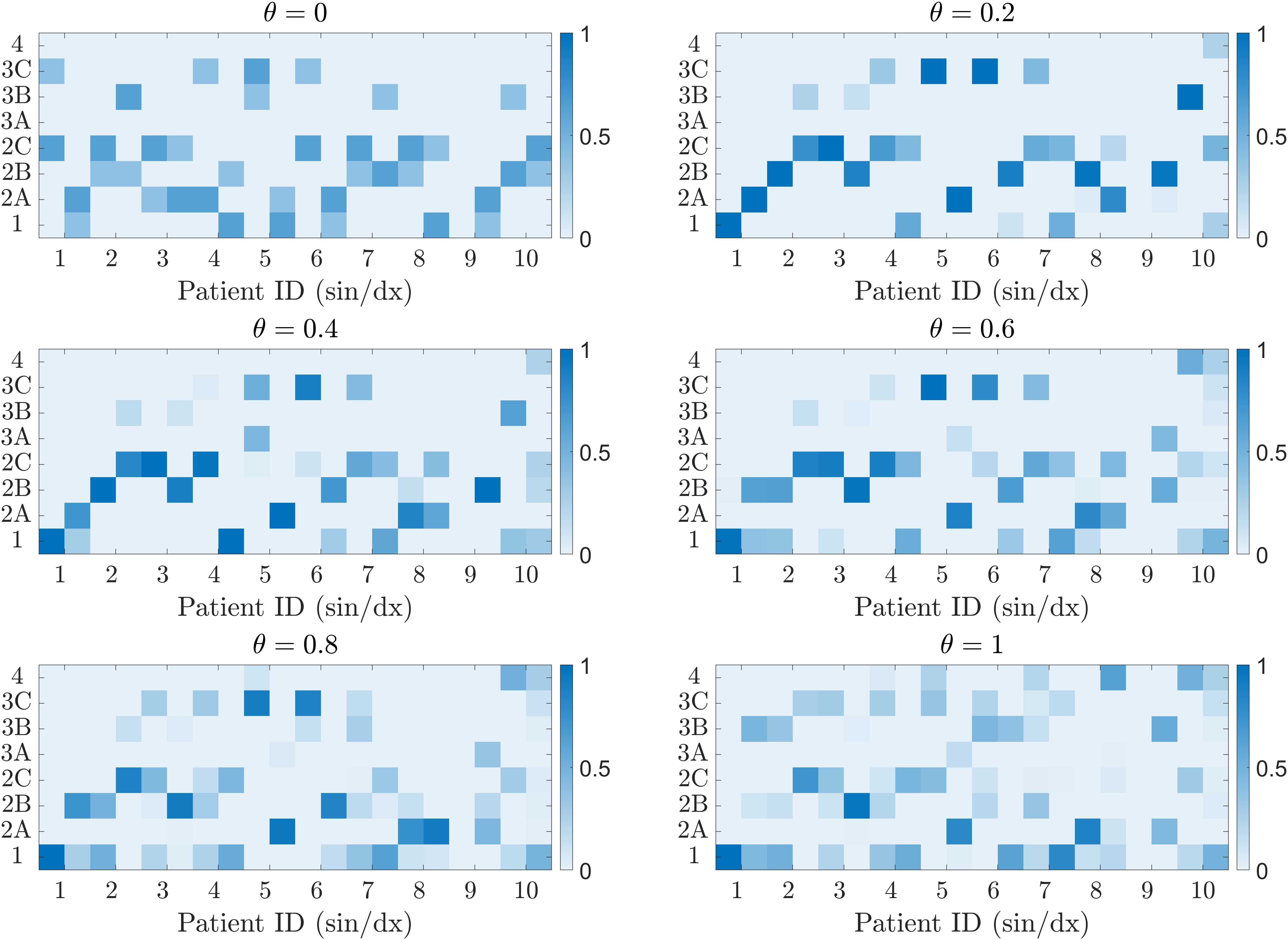}
    \caption{ The optimized current distributions for all ten patients, obtained using the LP formulation in~\eqref{eq:LP}, are shown for six different values of the constraint parameter $\theta$. The parameter $\theta$ determines the percentage of points in $\Omega_{\mathrm{c}}$ that must remain below the threshold $E_{\mathrm{th,c}}$. The color gradient from light blue ($\theta = 0$) to dark blue ($\theta = 1$) represents the proportion of the total applied current allocated to each contact.}
    \label{fig:LP_results_cohort}
\end{figure}

The results obtained from  solving the LP problem~\eqref{eq:LP} are presented in Fig.~\ref{fig:LP_results_cohort} for different relaxation levels $\theta$ and across the patient cohort.  
The suggested contact combinations are generally consistent over a range of $\theta$ values, but tend to blur out for high and low $\theta$ values.

\subsection{Mixed integer linear programming results}
\begin{figure}[htbp]
    \centering
    \includegraphics[width=0.8\linewidth]{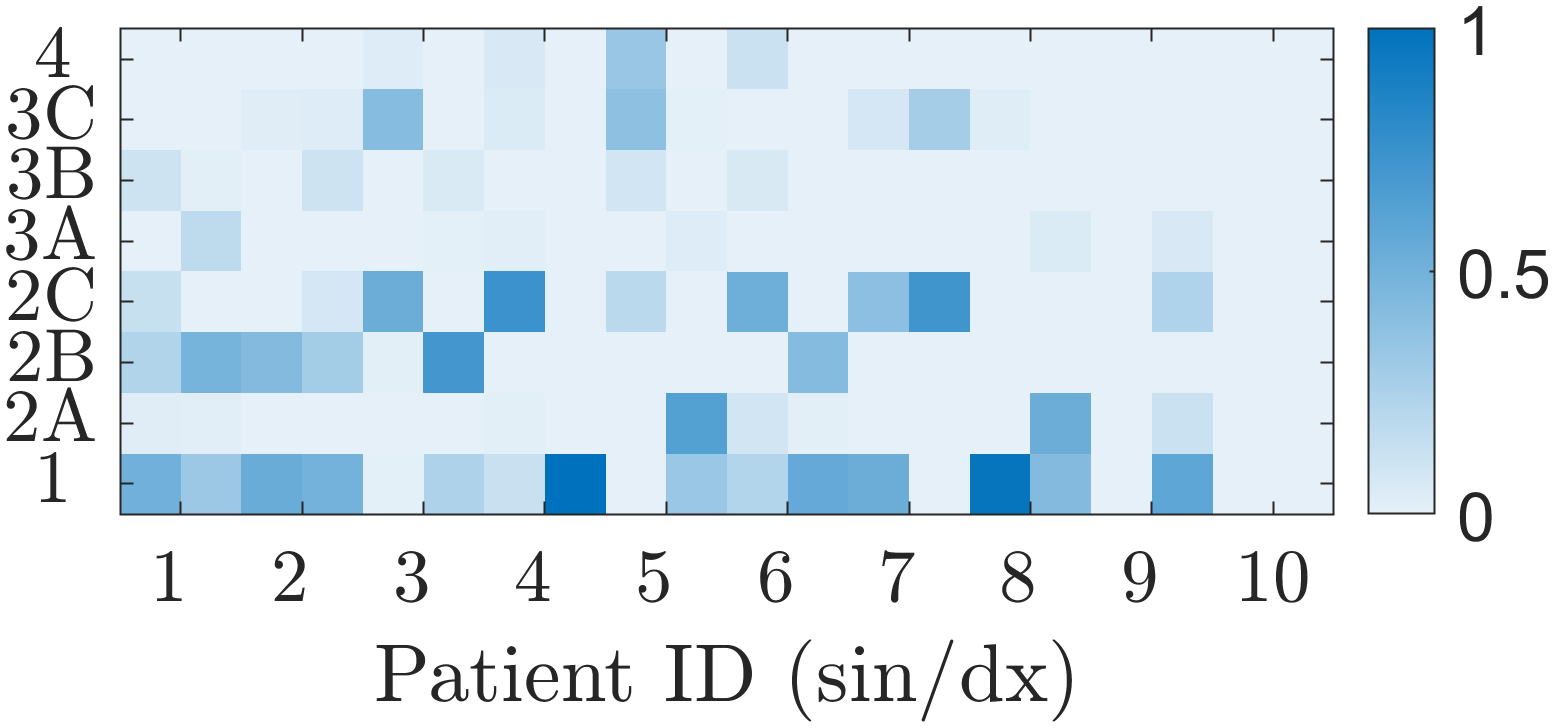}
    \caption{Optimized current distributions across all eight contacts computed from the MILP approach given in~\eqref{eq:MILP}. The values range from zero (light blue) to one (dark blue), representing the proportion of the total applied current allocated to each contact.}
    \label{fig:MILP_results}
\end{figure}

The optimized current distributions computed with the MILP approach are given in Fig.~\ref{fig:MILP_results}.

Notably, the MILP approach suggests zero amplitude in both leads of Patient~10. This is likely explainable by the poor lead placement relative to the atlas-based subdivisions of the STN in this patient. A more detailed analysis of this clinical case is provided in~\cite{Frigge2024PP}.

\subsection{Comparison}
\begin{figure}
    \centering
    \includegraphics[width=\linewidth]{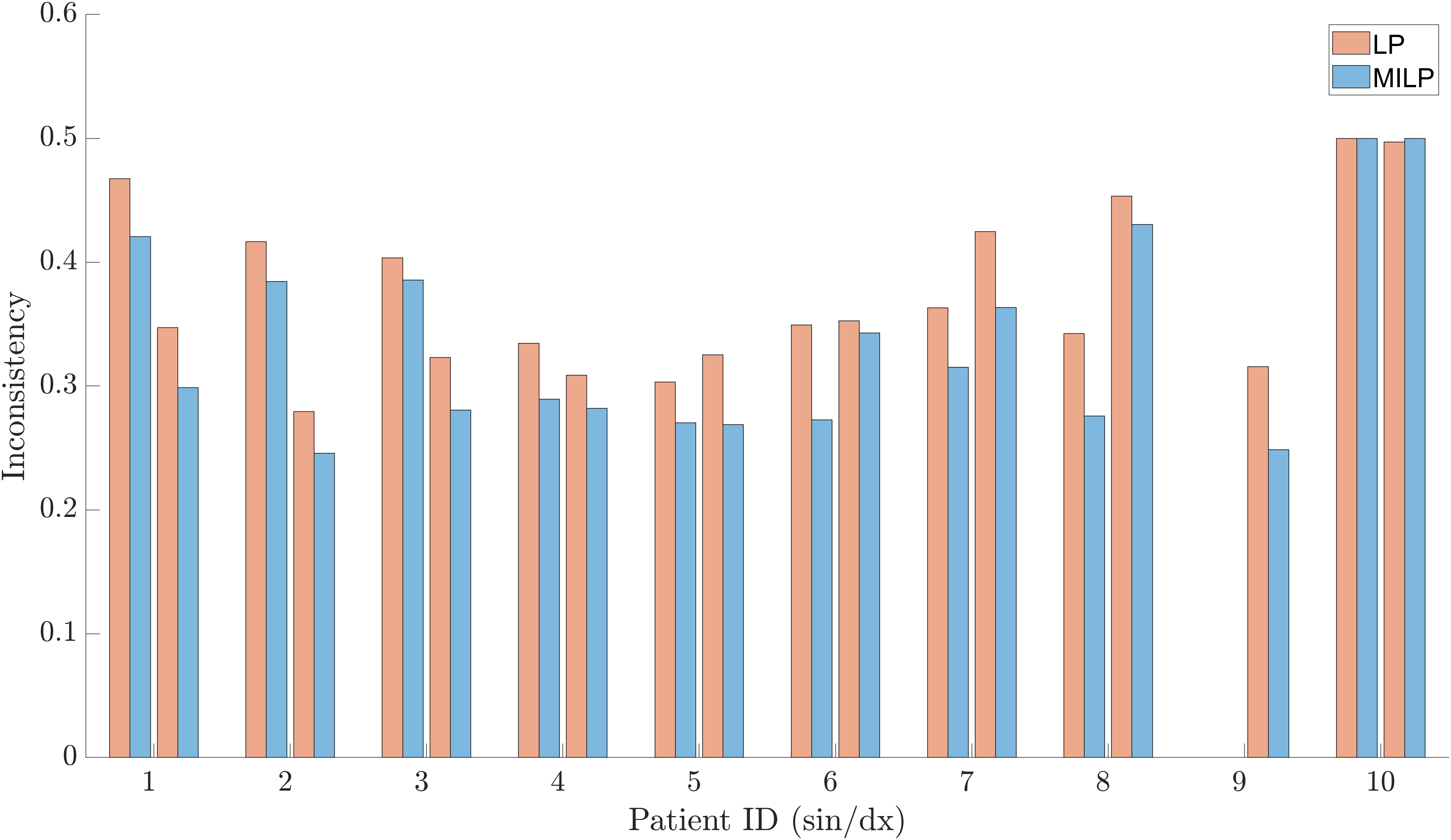}
    \caption{Inconsistency measure values~\eqref{eq:inconsistency} for both LP and MILP solutions across all patients. The MILP solution consistently achieves lower inconsistency, indicating better alignment with the desired activation profile. For Patient 10, both solutions exhibit high inconsistency, likely due to suboptimal lead positioning relative to the STN subdivisions.}
    \label{fig:inconsistency}
\end{figure}

Fig.~\ref{fig:inconsistency} illustrates the inconsistency metric values, as defined by~\eqref{eq:inconsistency}, for both the LP and MILP solutions across all patients. The MILP invariably achieves lower inconsistency, indicating a superior ability to match the desired activation profile. For Patient~10, both the LP and MILP solutions yield inconsistency values close to or equal to 0.5. As previously discussed, this is likely due to suboptimal lead positioning relative to the STN subdivisions, limiting the ability to achieve selective stimulation.

Results on the cohort level are compared using the Frobenius norm in~\eqref{eq:Frobenius}. In Tab.~\ref{tab:frobenius_comparison} the MILP and LP solution are compared to both each other and to the clinical settings presented in Fig.~\ref{fig:ClinicalSettings}. The Frobenius norm between clinical settings and the MILP solution is $3.52$. Between the clinical settings and the LP solution the value varies with the constraint parameter $\theta$. While the mean norm ($3.87 \pm 0.39$) is higher than the MILP norm, the LP outperforms the MILP at $\theta = 0$. However, on an individual patient level, the $\theta$ value that yields the closest match to the clinical vector of active contacts can vary significantly due to patient-specific sensitivity to stimulation. Although being off the primary focus of this study, it is worth noting that unconstrained optimization ($\theta=0$) often results in excessively high stimulation amplitudes, potentially failing to achieve an optimal balance between stimulation efficacy and target coverage.

When comparing MILP and LP solutions, the least difference occurs at $\theta=0.6$, thus providing the closest approximation of the MILP solution at the cohort level.

\begin{table}[h]
    \centering
    \caption{Comparison in Frobenius norm \eqref{eq:Frobenius} of the current distributions on the cohort level. The LP results (Fig.~\ref{fig:LP_results_cohort}) and MILP results (Fig.~\ref{fig:MILP_results}) are evaluated against each other and the clinical settings (Fig.~\ref{fig:ClinicalSettings}). }
    \label{tab:frobenius_comparison}
    \begin{tabular}{ccccccc}
        \toprule
         \multicolumn{6}{c}{\textbf{Clinical vs. MILP}} \\
        \cmidrule(lr){2-5} 
         \multicolumn{6}{c}{3.52} \\
        \midrule
         \multicolumn{6}{c}{\textbf{Clinical vs. LP}} \\
        \cmidrule(lr){2-5} 
         $\theta = 0.0$& $\theta = 0.2$ & $\theta = 0.4$ & $\theta = 0.6$ & $\theta = 0.8$ & $\theta = 1.0$  \\ 
        \cmidrule(lr){1-6}
         3.16 & 4.21 & 4.16 & 4.07 & 3.85 & 3.76\\
        \midrule
         \multicolumn{6}{c}{\textbf{MILP vs. LP}}  \\
        \cmidrule(lr){2-5} 
         $\theta = 0.0$& $\theta = 0.2$ & $\theta = 0.4$ & $\theta = 0.6$ & $\theta = 0.8$ & $\theta = 1.0$  \\ 
        \cmidrule(lr){1-6}
          3.23 & 3.31 & 3.06 & 2.74 & 2.76 & 2.91\\
        \bottomrule
    \end{tabular}
\end{table}

\begin{figure}
    \centering
    \includegraphics[width=\linewidth]{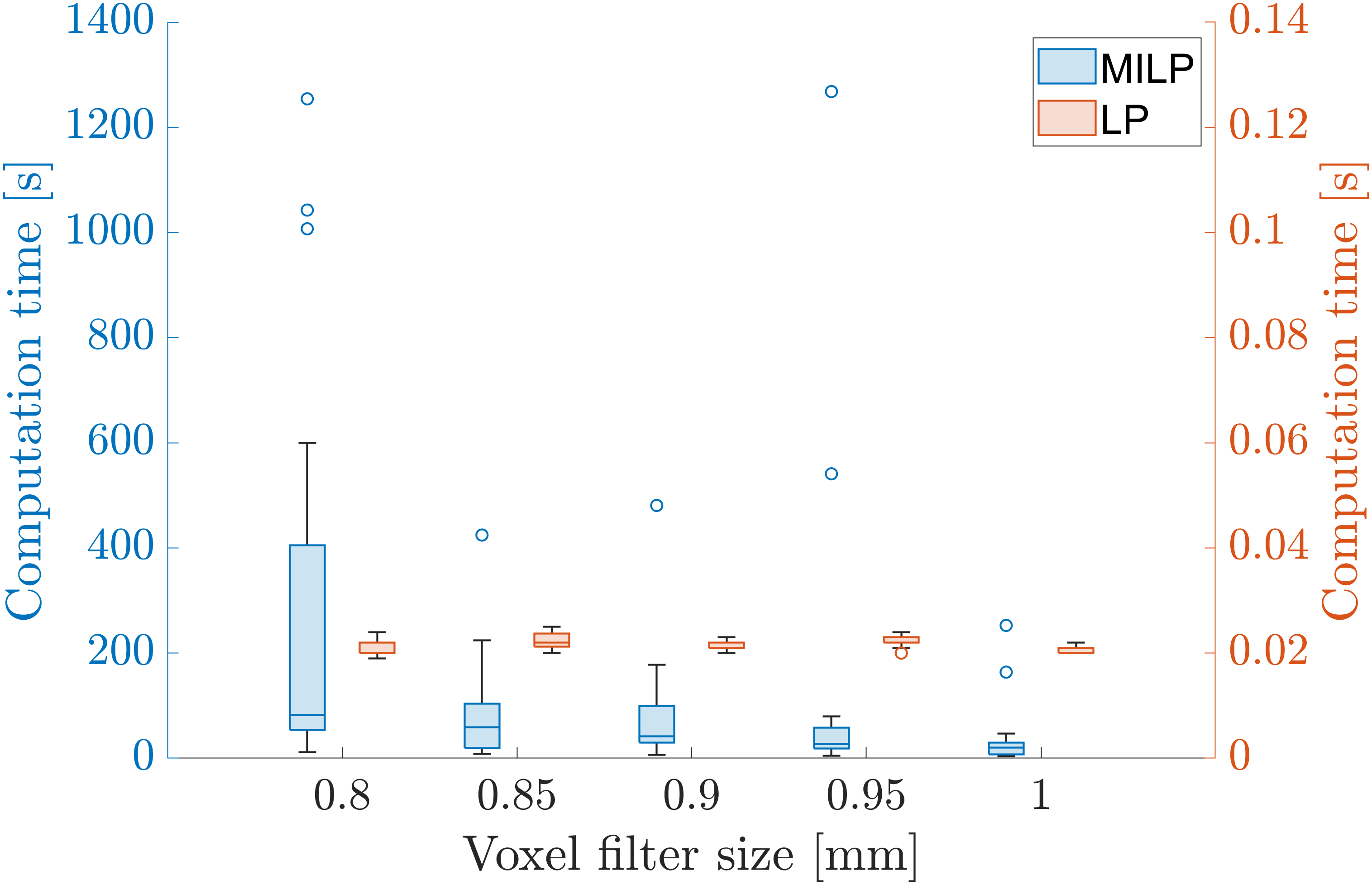}
    \caption{Runtimes for MILP and LP.  LP times are smaller by a factor $10,000$. Runtimes for MILP exceeded the time limit of $\SI{1500}{s}$ in three cases at $\SI{0.8}{mm}$ and in one case at $\SI{0.85}{mm}$. }
    \label{fig:Runtimes_milp}
\end{figure}

Fig.~\ref{fig:Runtimes_milp} presents the runtimes of the LP and the MILP algorithm, respectively, for different voxel filter sizes. Larger voxel filter sizes imply  smaller number of both target and constraint points, i.e. $N_{\mathrm{t}}$ and $N_{\mathrm{c}}$. As can be expected, the LP formulation is solved at a significantly lower computational cost, which is faster by a factor of 10,000. MILP runtimes exceeded the time limit of 1500 seconds in three cases at $\SI{0.8}{mm}$ and in one case at $\SI{0.85}{mm}$ voxel filter size.

\section{Discussion}

\subsection{Performance}
The results show that the MILP approach consistently outperformed the LP formulation demonstrating superior ability to achieve the desired activation profile. However, this comes at the cost of increased computational complexity. The LP formulation, on the other hand, is significantly more efficient and may be preferable when rapid optimization is required or when working with a high number of target and constraint points.

The LP solution is also sensitive to the choice of the relaxation parameter $\theta$, which determines the proportion of constraint points allowed to exceed the electric field threshold. While this behavior is expected, the optimal $\theta$ value for minimizing inconsistency is highly patient-specific and not known a priori.

\subsection{Comparison to clinical settings}
Comparing the optimization results to clinical DBS settings yielded ambiguous findings. This is not entirely surprising, as clinical programming is constrained by practical limitations. Only a limited number of stimulation settings can be tested in practice, meaning that the chosen parameters represent a feasible, but not necessarily optimal, configuration. Furthermore, uncertainties in lead localization and orientation reconstruction can influence which segmented contacts are deemed optimal, further complicating direct comparisons.

Despite the superiority of the MILP formulation in achieving a desired activation profile, the smallest difference between clinical settings and predicted settings were observed for the LP formulation with $\theta=0$. This is likely due to the choice of the desired activation profile, which may only insufficiently capture clinically relevant locations. 

Another important consideration is the variability in patient responses. Individual sensitivity to stimulation can fluctuate throughout the day and over time due to disease progression~\cite{Fasano2020}. As a result, while the optimization frameworks discussed in this paper provide a systematic approach to defining current distributions, the precise amplitude required for a given patient remains challenging to predict.

\subsection{Target selection}
This study defined target and constraint regions based on functional subdivisions of the STN from an atlas. However, research suggests that stimulation "sweet spots" within the STN or "sweet streamlines" along specific white matter tracts may offer more precise symptom relief~\cite{Dembek2019,Hollunder2024,Rajamani2024}. These regions are typically identified through large patient cohorts, but their effectiveness may vary significantly between individuals. Future work could explore more refined target definitions, potentially integrating patient-specific stimulation maps or functional imaging data to improve accuracy.
Additionally, the activation profile assumed in this study may be overly simplistic. In reality, therapeutic DBS likely modulates not only local structures but also connected white matter tracts. Future models could incorporate streamline activation or consider multiple, symptom-specific targets that need to be stimulated simultaneously~\cite{Hollunder2024,Rajamani2024}.

\subsection{Future work}
MILP achieves superior activation profiles, however its computational burden increases significantly with the number of target and constraint points, limiting its feasibility for large-scale applications. In contrast, LP provides a much faster solution, making it a practical choice when computational resources are limited.

To address this, future work could explore hybrid approaches that combine the strengths of LP and MILP—for example, using LP to generate an initial estimate on a larger set of points that MILP then refines in particular areas through more accurate sampling.  Additionally, reducing the number of constraint locations by incorporating more precise definitions of  target regions may allow for more complex activation patterns while maintaining computational feasibility.

Another modification could include the introduction of a constraint on the maximum number of active contacts. Reducing the number of simultaneously active contacts could lower impedance, thereby extending battery life - a relevant consideration for implanted pulse generators~\cite{Almeida2016}.

\section{Conclusions}
While the MILP optimization approach outperforms LP in achieving the desired activation profile in DBS programming, the predefined activation profile requires further refinement to better align with clinically applied settings. Additionally, future work should focus on comparing quantified patient response of the suggested configurations with clinically active settings to further assess and improve the utility of the optimization models.





\section*{ACKNOWLEDGMENT}
The computations were enabled by resources provided by the National Academic Infrastructure for Supercomputing in Sweden (NAISS) and the Swedish National Infrastructure for Computing (SNIC) at UPPMAX partially funded by the Swedish Research Council through grant agreements no. 2022-06725 and no. 2018-05973. Uppsala University Hospital. Additionally, the authors would like to thank Elena Jiltsova and Dag Nyholm at Uppsala University Hospital for providing the clinical data.


\bibliographystyle{ieeetr}
\bibliography{references}

\end{document}